\documentclass[twocolumn,amsmath,amssymb,prl,footinbib,10pt,superscriptaddress]{revtex4-1}
\usepackage{ graphicx, float,amsmath, amsmath, amssymb}
\usepackage[export]{adjustbox}
\newcommand{\qsubrm}[2]{{#1}_{\scriptscriptstyle{\textrm{#2}}}}
\newcommand{\qsuprm}[2]{{#1}^{\scriptscriptstyle\textrm{#2}}}
\def\be{\begin{equation}}
\def\ee{\end{equation}}
\def\bea{\begin{eqnarray}}
\def\eea{\end{eqnarray}}
\def\bse{\begin{subequations}}
\def\ese{\end{subequations}}

\usepackage[breaklinks, colorlinks,linkcolor=black, citecolor=black,urlcolor = black]{hyperref}
\usepackage[labelsep=period]{caption}
\begin{document}
\title{Observational Exclusion of a Consistent Quantum Cosmology Scenario}
\author{Boris Bolliet}
\email[Corresponding author: ]{boris.bolliet@lpsc.in2p3.fr}
\affiliation{%
Laboratoire de Physique Subatomique et de Cosmologie, Universit\'e Grenoble-Alpes, CNRS/IN2P3\\
53, avenue des Martyrs, 38026 Grenoble cedex, France
}
\author{Aur\'elien Barrau}%
\affiliation{%
Laboratoire de Physique Subatomique et de Cosmologie, Universit\'e Grenoble-Alpes, CNRS/IN2P3\\
53, avenue des Martyrs, 38026 Grenoble cedex, France
}
\author{Julien Grain}
\affiliation{%
CNRS,  Institut d'Astrophysique Spatiale, UMR8617, Orsay, France, F-91405}
\affiliation{%
 Universit\'e Paris-Sud 11, Orsay, France, F-91405}
\author{Susanne Schander}%
\affiliation{%
Laboratoire de Physique Subatomique et de Cosmologie, Universit\'e Grenoble-Alpes, CNRS/IN2P3\\
53, avenue des Martyrs, 38026 Grenoble cedex, France
}
\date{\today}
\begin{abstract} 
It is often argued that inflation erases all the information about what took place before it started. Quantum gravity, relevant in the Planck era, seems therefore mostly impossible to probe with cosmological observations. In general, only very \textit{ad hoc} scenarios or hyper fine-tuned initial conditions can lead to observationally testable theories. Here we  consider a well-defined and well motivated candidate quantum cosmology model that predicts inflation. Using the most recent observational constraints on the cosmic microwave background B modes, we show that the model is excluded for all its parameter space, without any tuning.  Some important consequences are drawn for the \textit{deformed algebra approach} to loop quantum cosmology. We emphasize that neither loop quantum cosmology in general nor loop quantum gravity are disfavored by this study but their falsifiability is established.
 \end{abstract}
\maketitle
\begin{figure}[H]
\includegraphics[width=85mm,left]{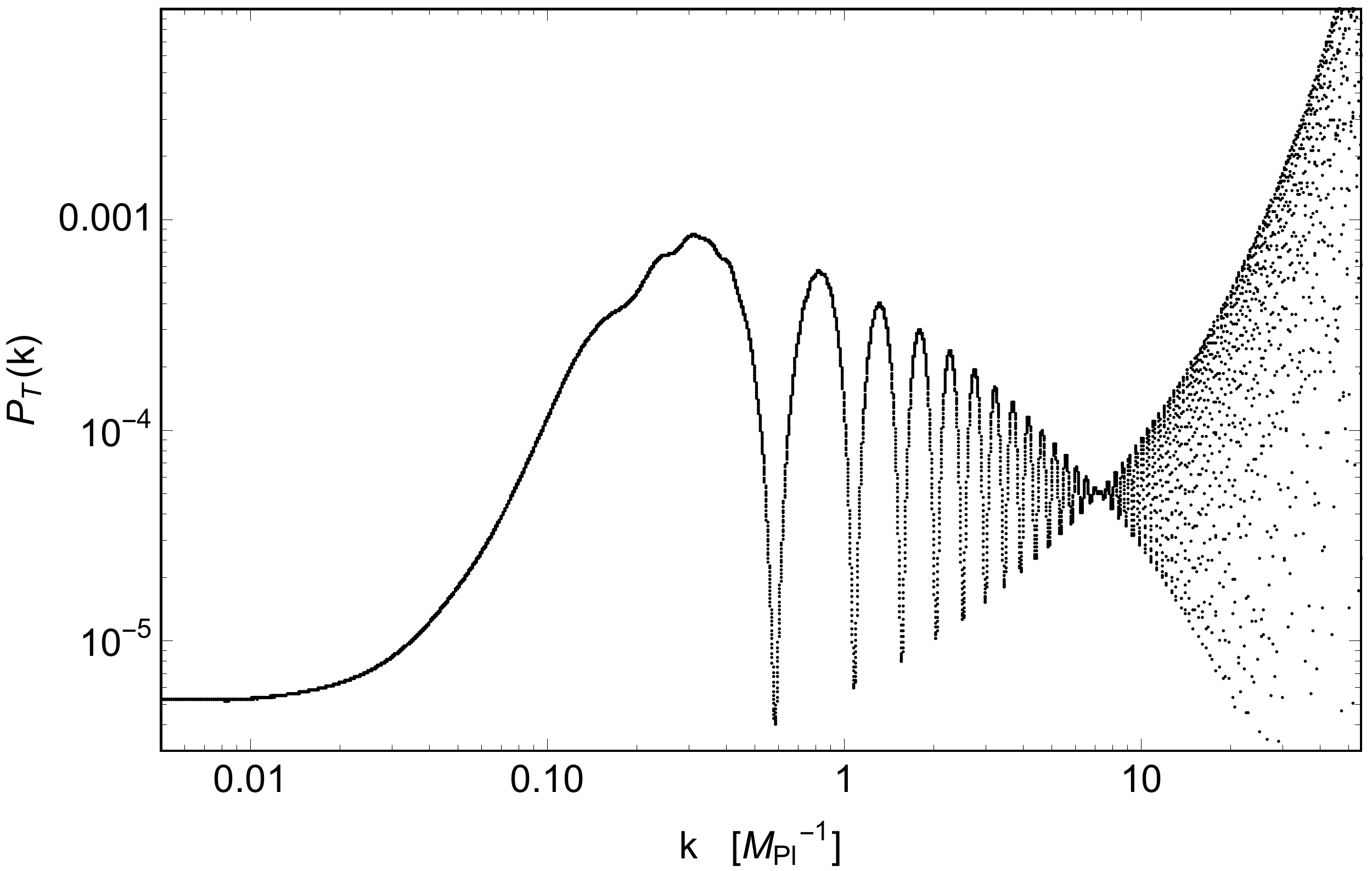}
\caption{Tensor primordial power spectrum predicted by the \textit{deformed algebra approach} to cosmological perturbation in loop quantum cosmology. The data points were obtained from a numerical simulation (with mass $m=1.2\times10^{-3}\,\qsubrm{m}{Pl}$ to improve the numerical stability on the exponential part).}
\label{fig:Pk}
\end{figure}
\begin{figure}[H]
\includegraphics[width=95mm,center]{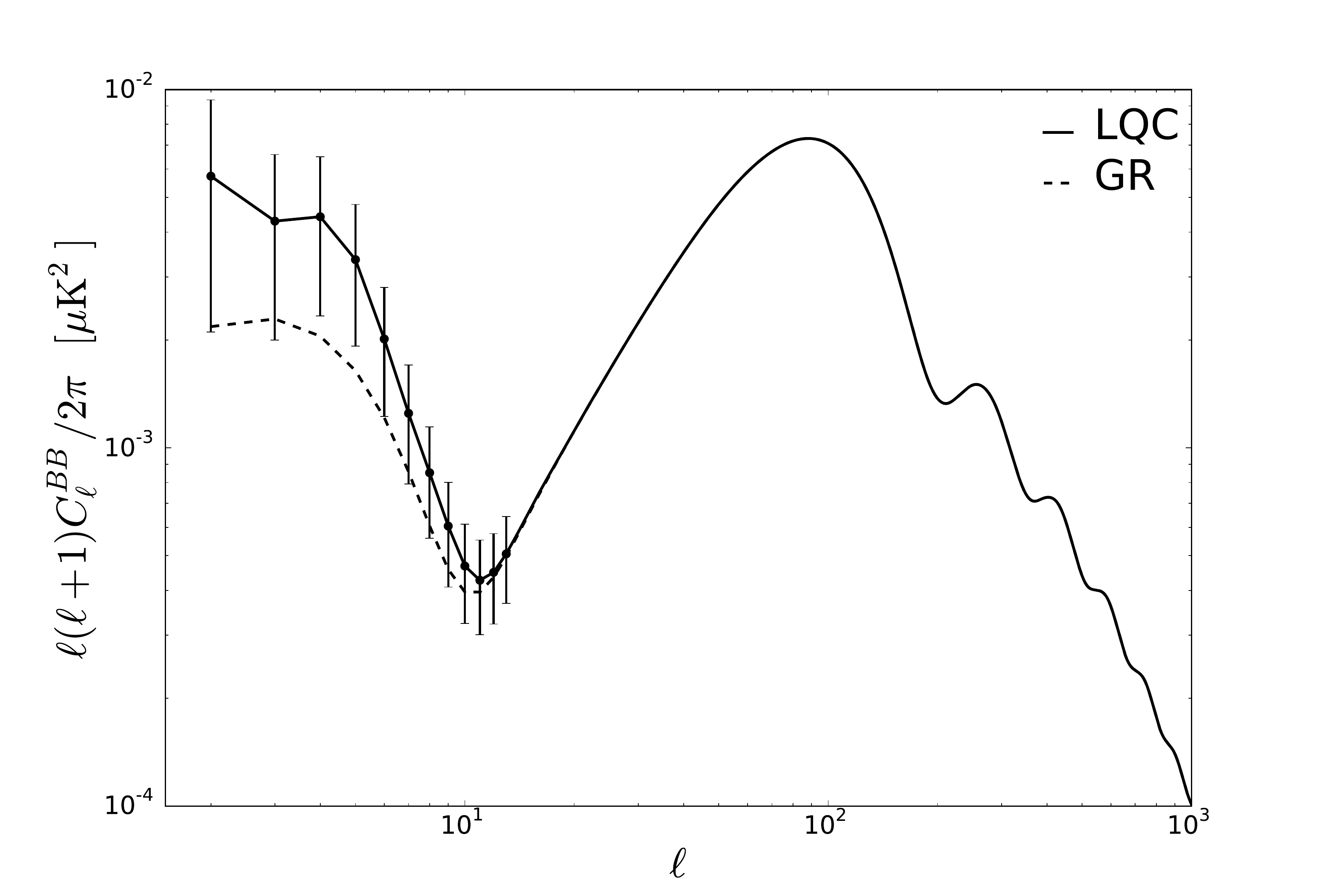}
\caption{Cosmic Microwave Background B modes angular power spectra (wihtout lensing) obtained with CLASS \cite{CLASS}, using the best fit parameters for TT+LowP+Lensing in Table 4 of \cite{PCP2015}, for the primordial spectrum obtained in the \textit{deformed algebra approach} with the exponential rise arbitrary removed and replaced by the red-tilted spectrum as predicted in standard inflation with
$k_\star=0.002\,\mathrm{Mpc}^{-1}$, $\qsubrm{A}{s}=2.139\times10^{-9}$ and $\qsubrm{r}{002}=0.114$. The signal expected in standard general relativity corresponds to the dashed line.
The error bars, shown for the first twelve multipoles, correspond to the cosmic variance. The specific oscillations in the primordial power spectrum are clearly washed out.}
\label{fig:BB}
\end{figure}

\textit{Introduction}.---This Letter aims at giving a concrete example of a fully consistent quantum cosmology scenario with general relativity (GR) as its low-energy limit and leading to a standard phase of inflationary expansion of the Universe that is excluded by current experimental data. Although the considered model belongs to the loop quantum cosmology (LQC) framework, we emphasize from the beginning that the claim is not that loop quantum gravity (LQG) or LQC is excluded. The other way round: the fact that some specific settings within LQC are excluded demonstrates that the theory can fill the bridge between calculations and observations, which makes it an especially appealing quantum gravity proposal.\\
\indent LQG is a non-perturbative  background-independent quantization of GR  \cite{lqg1,lqg2,lqg3,lqg4,lqg5,lqg6,lqg7,lqg9,lqg10} that relies on the Ashtekar variables, an SU(2)-valued connection and its conjugate densitized triads. The quantization is performed over the holonomies of the connections and the fluxes of the densitized triads. 
Important questions, in particular regarding the continuum limit of LQG,  remain opened but important progresses have been achieved recently, see {\it e.g.} \cite{barrett,han}. \\
\indent LQC is a symmetry reduced version of LQG using cosmological symmetries. In LQC, the big bang is replaced by a big bounce due to repulsive quantum gravity effects close to the Planck density \cite{lqc1,lqc2,lqc3,lqc4,lqc5,lqc6,lqc7,lqc9,lqc10,lqc11,lqc12}. It is however important to underline that LQC has not yet been rigorously derived from LQG and remains an attempt to use LQG-like methods in the cosmological sector. \\\indent Although there is a general agreement on the background dynamics in LQC (modulo some issues on the best motivated initial conditions),
there are different ways to  implement LQG ideas at the level of cosmological perturbations. The most popular models are the {\it dressed metric approach}, the {\it hybrid quantization approach} and the {\it deformed algebra approach}. The {\it dressed metric} hypothesis \cite{Agullo1,Agullo2,Agullo3} accounts for quantum fields propagating on a quantum background but lacks a proof of consistency taking into account the subtle gauge issues in gravity \cite{barrett,han}. 
The {\it hybrid quantization} formalism \cite{merce} nicely takes backreaction effects  into account, but remains at a very early stage of development. In this work we focus on the {\it deformed algebra approach}  \cite{eucl3} which is probably the most developed one and has generated a very large number of articles (see {\it e.g.} \cite{Bojowald:2011aa} and references therein).\\
\indent \textit{The deformed algebra approach}.---The fact that holonomies of the connections are the basic LQG variables can be accounted for, at the effective level, by the \textit{standard holonomy correction} in the Hamiltonian constraint which consists in replacing the mean Ashtekar connection by a pseudo periodic function depending on the coordinate size of a loop (see {\it e.g.} \cite{bojo1,bojo2,bojo3} for seminal articles).  
The crucial point of the {\it deformed algebra approach} is to ensure that the resulting Poisson algebra remains consistent, so that Poisson brackets between quantum corrected constraints are proportional to a quantum corrected constraint. This algebraic structure has been derived for vector modes \cite{tom1}, scalar modes \cite{tom2} and shown to be consistent for tensor modes \cite{eucl2}. Although requiring quite a lot of algebra, the main result is surprisingly simple and impressive in the way that there exists a \textit{unique} solution to the anomaly freedom problem, which is far from being trivial. Furthermore, this procedure determines the lattice refinement scheme to be precisely the desired one, that is the so-called $\bar{\mu}$-scheme  \cite{Ashtekar:2006wn}.
The resulting {\it anomaly free} algebra reads
\begin{eqnarray}
\left\{D[M^a],D [N^a]\right\} &=& D[M^b\partial_b N^a-N^b\partial_b M^a], \nonumber \\
\left\{D[M^a],S[N]\right\} &=& S[M^a\partial_b N-N\partial_a M^a],  \\
\left\{S[M],S[N]\right\} &=& \Omega
D\left[q^{ab}(M\partial_bN-N\partial_bM)\right], \nonumber
\end{eqnarray}
where $N^{a}(M^a)$ and $N(M)$ are the shift and lapse functions, $q^{ab}$ is the spatial metric, $D$ and $S$ are the holonomy-corrected diffeomorphism and hamiltonian constraints, and $\Omega \equiv1-2\rho/\qsubrm{\rho}{B}$ with $\rho$ the density of the universe and $\qsubrm{\rho}{B}$ the critical density (close to the Planck density) which encodes deviations from standard GR.
The algebra is elegant and simple. Furthermore it leads to a signature change close to the bounce which is somehow reminiscent of the Hartle-Hawking proposal \cite{HHP}. 
 When $\rho<(\qsubrm{\rho}{B}/2)$ the spacetime geometry is Lorentzian, but when $\rho>(\qsubrm{\rho}{B}/2)$, in the vicinity of the bounce, $\Omega$ becomes negative and the spacetime geometry becomes Euclidean. Strikingly, this effect has been found independently, still in LQC, in \cite{bp} and \cite{ed}, the latter approach relying on a different approach based on ``patches of universe" evolving independently in the longitudinal gauge.\\ 
\indent \textit{Model and assumptions}.---Here, we shall investigate the observational consequences of this effective signature change for cosmological perturbations near the big bounce, under some basic hypotheses: (i) we assume the universe to be filled with a massive scalar field, that is with a potential $V(\phi)=m^2\phi^2/2$, (ii) we restrict ourselves to spatially flat Friedmann-Lemaitre-Robertson-Walker cosmologies, (iii) we assume that initial conditions should be set in the remote past of the classical contracting branch of the universe for both the background and the perturbations, as expected from a truly causal evolution, (iv) we assume that initially the quantum field describing the metric perturbations is in the Minkowski vacuum state, which is well defined and non-ambiguous, (v) as argued in \cite{Agullo1,Agullo2,Agullo3}, we assume that the calculated perturbation power spectra make sense up to arbitrary small scales as long as the energy of the perturbations remains small when compared to the energy of the background, (vi) we do not consider backreaction effects. Each of these hypotheses can obviously be questioned. However, quite obviously too, each of them constitutes the most ``natural", usual, and simple choice. 
\indent The metric perturbations are evolved from their initial vacuum state, in the contracting universe, toward some specific state, in the expanding branch, that can be computed numerically, allowing one to obtain the primordial power spectrum used as an input for cosmic microwave background (CMB) phenomenology.\\
\indent The primordial tensor and scalar power spectra were derived in \cite{lcbg}  and \cite{susanne} respectively.  Here, we focus on tensor modes that are better controlled (and are enough for our conclusion). The key point for phenomenology that we shall now investigate into more details is the duration of inflation since it allows one to convert the comoving wavenumbers, used in these studies, into physical scales probed by CMB experiments. Although, as demonstrated in \cite{bl}, there exists a highly favored number of e-folds of inflation in this model, $\qsubrm{N}{tot}\simeq140$, associated with a  value of the scalar field at the bounce of about $\qsubrm{\phi}{B}\simeq 2.6\,\qsubrm{m}{Pl}$ with $\qsubrm{m}{Pl}\equiv 1/\sqrt{G}$, the conclusion we shall reach in the following remains true even for fine-tuned initial background conditions leading to a higher or smaller number of e-folds.\\ 
\indent \textit{Observational constraints}.---For CMB phenomenology, the scales of interest range between $\qsubrm{k}{min}=10^{-6}\,\mathrm{Mpc}^{-1}$ and $\qsubrm{k}{max}=1\,\mathrm{Mpc}^{-1}$. This range is referred to as the \textit{observable window} of wavenumbers. The energy density at the bounce is associated to a wavenumber
\be
 \qsubrm{k}{B}\equiv\qsubrm{a}{B}\sqrt{\qsubrm{\rho}{B}}\qsubrm{M}{Pl}^{-1},
 \ee
where $\qsubrm{\rho}{B}$ has dimensions of $\qsubrm{M}{Pl}^{4}$, where this time $\qsubrm{M}{Pl}\equiv\qsubrm{m}{Pl}/\sqrt{8\pi}$ denotes the reduced Planck mass. 
The evolution of the amplitude of the fluctuations in the early universe depends on their size when compared to the horizon size (weighted by the $\Omega$ factor). Horizon-crossing is defined as the time when the wavelength of the considered perturbation equals the Hubble horizon. In particular, for the pivot scale $k_\star$, used to parametrize the primordial power spectra, this reads $k_\star=a_\star H_\star$, where $a_\star$ and $H_\star$ are the scale factor and the Hubble parameter at horizon-crossing during inflation. 
Clearly, possible footprints of LQC effects in the angular power spectrum of CMB anisotropies depend on the relative values of $k_\star$ and $\qsubrm{k}{B}$. This motivates the definition of a dimensionless function: 
\be
n(\qsubrm{\rho}{B},\qsubrm{\phi}{B})\equiv\ln (k_\star/\qsubrm{k}{B}).
\ee 
The dependence upon the value of the scalar field at the bounce, $\qsubrm{\phi}{B}$, will appear explicitly in the following. \\
\indent As shown in Fig.\,\ref{fig:Pk}, the main characteristics of the spectrum we are considering in this study are \cite{lcbg}: (i) scale invariance for the infrared (IR) scales, \textit{i.e.} $k\ll\qsubrm{k}{B}$, (ii) oscillations for the intermediate scales, \textit{i.e.} $k \sim \qsubrm{k}{B}$, and (iii) an exponential growth for  the ultraviolet (UV) scales, \textit{i.e.} $k\gg \qsubrm{k}{B}$. \\
\indent Let us now describe in more details how the scales affected by LQC effects compare to the present \textit{observable window} of wavenumbers. The condition $\qsubrm{k}{B}\sim\qsubrm{k}{max}$ reads $n(\qsubrm{\rho}{B},\qsubrm{\phi}{B})\simeq-6.2$ when $\qsubrm{k}{max}$ is replaced with the numerical value given above. Our first goal is to analyze how such a condition can be fulfilled depending on the values of $\qsubrm{\rho}{B}$ and $\qsubrm{\phi}{B}$. Expanding the ratio $(k_\star/\qsubrm{k}{B})$ over the cosmic history, from the bounce until horizon-crossing, one gets
\be
n(\qsubrm{\rho}{B},\qsubrm{\phi}{B})=\qsubrm{N}{tot}-N_\star+\qsubrm{N}{B}+\tfrac{1}{2} \ln (V_\star/3\qsubrm{\rho}{B}),\label{eq:n}
\ee
where $V_\star$ is the potential energy of the scalar field at horizon-crossing, $\qsubrm{N}{B}$ is the number of e-folds between the bounce and the start of inflation, $\qsubrm{N}{tot}$ is the total number of e-folds of the inflationary phase, and $N_\star$ is the number of e-folds of observable inflation, \textit{i.e.} from horizon-crossing until the end of inflation. This number can in turn be calculated in terms of quantities related to the post-inflationary evolution of the universe as \cite{Liddle:1993fq,Ade:2015lrj} 
\bea
N_\star&=&-\ln({k_\star}/\qsubrm{k}{0})+\tfrac{1}{4}\ln(\Omega_\gamma \qsubrm{M}{pl}^2/3\qsubrm{H}{0}^2)-\tfrac{1}{12}\ln\qsubrm{g}{th}\nonumber\\
&&+\tfrac{1}{12}\ln(\qsubrm{\rho}{th}/\qsubrm{\rho}{end})+\tfrac{1}{2}\ln(V_\star/\sqrt{\qsubrm{\rho}{end}}\qsubrm{M}{pl}^2).\label{eq:nstar}
\eea
In this formula, some parameters are known:  $\qsubrm{k}{0}\equiv\qsubrm{H}{0}/c$ with $\qsubrm{H}{0}= (67.31\pm0.96) \,\mathrm{km\cdot s^{-1}\cdot Mpc^{-1}}$, the pivot scale is $k_\star=0.002\,\mathrm{Mpc}^{-1}$ by convention and the present radiation density of the universe is $\Omega_\gamma=5.45\times10^{-5}$. Moreover one can safely assume $\qsubrm{g}{th}\sim10^3$, accounting for the creation of new degrees of freedom during reheating. There remains three unknowns: (i) the potential energy at horizon-crossing, $V_\star=\tfrac{3}{2}\pi^2\qsubrm{A}{s}r_\star \qsubrm{M}{Pl}^4$, for which there exists an upper bound due to observational constraints on both the amplitude of the scalar primordial power spectrum, $\qsubrm{A}{s}$, and the tensor-to-scalar ratio, $r_\star$, (ii) the energy scale of reheating, $\qsubrm{\rho}{th}$, (iii) the energy density at the end of inflation, $\qsubrm{\rho}{end}$, which should lie between $\qsubrm{\rho}{th}$ and $V_\star$, and can be expressed as $\qsubrm{\rho}{end}\simeq m^2 \qsubrm{M}{Pl}^2$ for a massive scalar field. However, as discussed in \cite{Liddle2003},  most reasonable assumptions  (if not all) over the different energy scales lead to  take $N_\star$ between $50$ and $60$. We chose $N_\star=60$ in the following to illustrate the procedure (the other extreme choice would not change the conclusion nor the following numerical estimates by more than ten percents). This sets all the unknowns since $m$ can directly be expressed in terms of $N_\star$, leading to $m\simeq1.2\times10^{-6}\,\qsubrm{m}{Pl}$ for $N_\star=60$. The value of the mass can receive small LQC corrections \cite{Agullo2015} that play no role in our conclusion. Furthermore, now looking at (\ref{eq:n}) and using the analytical results of \cite{Bolliet:2015bka} one finds
$\qsubrm{N}{B}=\tfrac{1}{3}\ln\Gamma$, where $\Gamma\equiv \sqrt{3\qsubrm{\rho}{B}}/(m\qsubrm{M}{Pl})$, and $\qsubrm{N}{tot}=\tfrac{1}{4}(\qsubrm{\phi}{i}/\qsubrm{M}{Pl})^2-\tfrac{1}{2}$, with
\bea
\qsubrm{\phi}{i}&=&\qsubrm{\phi}{B}+\sqrt{\tfrac{2}{3}}\mathrm{Arcsinh}(\Gamma\sqrt{{2}/{\mathrm{W}\left(z\right)}})\label{eq:phii}
\eea
the scalar field at the start of slow-roll inflation, where the argument of the Lambert $\mathrm{W}$-function is 
\begin{equation}
z\equiv8\Gamma^{2}\exp(\sqrt{6}{\qsubrm{\phi}{B}}/\qsubrm{M}{Pl}). \label{eq:f}
\end{equation}
\indent \textit{Results}.---In LQC, the critical energy density, $\qsubrm{\rho}{B}$, is related to the fundamental parameter of LQG, $\gamma$, the  Barbero-Immirzi parameter, as
$
\qsubrm{\rho}{B}=2\sqrt{3}\qsubrm{M}{Pl}^4/\gamma^3  
$ 
\cite{Agullo3} (we do not, here, go into the subtlety of the area gap
definition \cite{Ashtekar:2015lla}). 
The standard value, $\gamma=0.2375$, sets $\qsubrm{\rho}{B}$ to $0.41\,\qsubrm{m}{Pl}^4$. In fact, within the slow-roll approximation and for a massive scalar field, all quantities in (\ref{eq:n}) and (\ref{eq:nstar}) can be expressed in terms of $\qsubrm{\phi}{B}$, once $N_\star$ has been fixed. It is then easy to obtain the value $\qsubrm{\phi}{B}^{\scriptscriptstyle{+}}$ such that $n(\qsubrm{\rho}{B},\qsubrm{\phi}{B}^{\scriptscriptstyle{+}})=-6.2$. With the previous numerical values we find $\qsubrm{\phi}{B}^{\scriptscriptstyle{+}}\simeq1.1\,\qsubrm{m}{Pl}$ corresponding to $\qsubrm{k}{B}=1\,\mathrm{Mpc}^{-1}$. This is far below the favored value of the scalar field at the bounce, $\qsubrm{\phi}{B}\simeq2.6\,\qsubrm{m}{Pl}$, which leads to a physical wavenumber $\qsubrm{k}{B}$  much smaller than $k_\star$. More precisely, the favored value of the field leads to $\qsubrm{k}{B}\simeq10^{-37}k_\star$. In other words, the scales of the primordial power spectrum that are probed by present measurements of the CMB anisotropies correspond to scales that were at the bounce much smaller than the characteristic scale of the bounce. This corresponds to the deep UV regime of the primordial power spectra presented in Fig.\,\ref{fig:Pk} \cite{lcbg,bl,Bolliet:2015bka}, that is the one clearly excluded by data: the exponential growth of the amplitude of the power spectrum at these scales is ruled out by the CMB upper bound on B-modes ($\qsubrm{r}{0.002}<0.114$). Obviously, backreaction should be taken into account for a detailed prediction, but the general trend, that is perturbations becoming huge due to the real exponential factor associated with the change of signature of the metric, will anyway contradict the stringent upper bound coming from current observations.
\indent With the natural measure defined in \cite{bl}, the probability to escape this exclusion by having $\qsubrm{k}{max}<\qsubrm{k}{B}$, that is $\qsubrm{\phi}{B}< \qsubrm{\phi}{B}^{\scriptscriptstyle{+}}$ with $\qsubrm{\phi}{B}^{\scriptscriptstyle{+}}\simeq1.1\,\qsubrm{m}{Pl}$, is less than $10^{-8}$. One could still focus on this specific case ($\qsubrm{\phi}{B}< \qsubrm{\phi}{B}^{\scriptscriptstyle{+}}$) by fine-tuning initial conditions for the background to this aim.  (This is what is usually done in phenomenological studies of the \textit{dressed metric approach} \cite{AS2011}, requiring in addition $\qsubrm{\phi}{B}>0.8\,\qsubrm{m}{Pl}$ so that the \textit{observable window} falls just on the interesting part of the spectrum.) The key point we want to underline is that such a fine-tunning would not save the model. The \textit{observable window} can fall in the ``low $k$" part of the spectrum only if there is less inflation. The duration of inflation would need to be very close  to its minimal value, $\qsubrm{N}{tot}\simeq60$. However, a detailed numerical study shows that to achieve such a small amount of inflation the universe must go through a long phase of \textit{deflation} (exponential decrease of the scale factor) before the bounce. This has a direct consequence on the primordial power spectrum:  due to the specific dynamics of deflation, the nearly scale-invariant IR part of the spectrum is drastically amplified. Indeed, 
the equation of propagation in conformal time for a tensor mode $v_k$ during deflation reduces to
\be
v_k^{\prime\prime}+(k^2-2\mathcal{H}^2)v_k=0,
\ee
where $\mathcal{H}$ is the conformal Hubble parameter. This equation is clearly unstable for the IR modes with $k<\sqrt{2}\mathcal{H}$.
So even if the shape of the portion of the spectrum which would be observable is correct, its normalization would exceed the observational upper bound  by orders of magnitude and the model would remain excluded. This has been numerically checked in details \footnote{For initial conditions such that $\qsubrm{N}{tot}\simeq60$ (corresponding to $\qsubrm{\phi}{B}\simeq5\,\qsubrm{M}{Pl}$) the excess in the amplitude of the primordial tensor modes is about fourteen orders of magnitudes higher than the observational upper bound given by the tensor-to-scalar ratio.}. \\ 
\indent Finally, it is important to mention that, even if the exponential growth of the spectrum is arbitrarily removed, and initial conditions chosen so that the \textit{observable window} falls exactly on the oscillatory part of the spectrum, the model would anyway hardly lead to a specific observational signature. This can be concluded from Fig.\,\ref{fig:BB} where the CMB $\qsuprm{C_\ell}{BB}$ were explicitly calculated. The specific oscillatory behavior predicted for the primordial tensor power spectrum in LQC, which is the key prediction here, is smeared out by the cosmic evolution. The LQC spectrum would remain mostly undistinguishable from the standard prediction (GR) due to the cosmic variance. 
\indent \textit{Discussion and conclusion}.---There has recently been a number of new results on primordial perturbations within the \textit{dressed metric approach}  \cite{Bonga2015,Gupt2015,Agullo2015}. The associated  primordial power spectra  were compared to those from the \textit{deformed algebra approach}  in \cite{Bolliet:2015bka}. They share the same features in the IR regime (a slightly red-tilted spectrum) and at intermediate scales (oscillatory behavior), but their UV regimes are very different as there is no exponential amplification in the \textit{dressed metric approach}. It is also worth emphasizing that the authors of the latter approach usually set initial conditions at the bounce, therefore avoiding by construction the effects of \textit{deflation}.\\
\indent Our main conclusion is that although the quantum cosmology model that is considered in this work is well-defined, well-motivated, has the standard Friedmann equation as its low-density limit and, even more importantly, leads to the required amount of inflation, it is excluded by current data. This illustrates with a concrete example that the usual statement claiming that ``whatever happens before inflation cannot be probed" is incorrect. Cosmological tests of quantum gravity are now possible, even with mainstream models without any tuning of the parameters. However it is   important to underline that only a very specific version of LQC is excluded: a universe filled with a massive scalar field, treated in the {\it deformed algebra approach}, with initial conditions set in the remote past before the bounce, no backreation, no anisotropies and no cutoff scale. This is, in itself, a substantial result to establish loop quantum cosmology as a predictive theory. \\

The work of B.B.  was supported by a grant from ENS Lyon. 
\bibliography{refs}
 \end{document}